\documentclass{PoS}

\title{High-resolution radio imaging of the gamma-ray blazar candidate J1331+2932}

\ShortTitle{Gamma-ray blazar candidate J1331+2932}

\author{A. Gemes\thanks{The EVN is a joint facility of independent European, African, Asian, and North American radio astronomy institutes. Scientific results from data presented in this publication are derived from the following EVN project code: RSG08. This publication makes use of data products from the \emph{WISE}, which is a joint project of the University of
California, Los Angeles, and the Jet Propulsion Laboratory/California Institute of Technology, funded by the National Aeronautics and Space Administration. This project has received funding from the European Union's Horizon 2020 research and innovation programme under grant agreement No 730562 (RadioNet). We thank the Hungarian National Research, Development and Innovation Office (OTKA NN110333 and 2018-2.1.1-UK\_GYAK) for support. K\'EG was supported by the J\'anos Bolyai Research Scholarship of the Hungarian Academy of Sciences.}\,\,,$^a$ \speaker{K. \'E. Gab\'anyi},$^{bc}$ S. Frey,$^{c}$ T. An,$^{d}$ Z. Paragi,$^{e}$ and A. Mo\'or$^{c}$\\
\llap{$^a$}Trinity College, University of Cambridge, Cambridge, United Kingdom\\
\llap{$^b$}MTA-ELTE Extragalactic Astrophysics Research Group, Budapest, Hungary\\
\llap{$^c$}Konkoly Observatory, MTA Research Centre for Astronomy and Earth Sciences, Budapest, Hungary\\
\llap{$^d$}Shanghai Astronomical Observatory, Chinese Academy of Sciences, Shanghai, China\\
\llap{$^e$}Joint Institute for VLBI ERIC, Dwingeloo, The Netherlands\\
E-mail: \email{ag996@cam.ac.uk}, \email{krisztina.g@gmail.com}, \email{frey.sandor@csfk.mta.hu}, \email{antao@shao.ac.cn}, \email{zparagi@jive.eu}, \email{moor@konkoly.hu} }

\abstract{Active galactic nuclei are the most luminous persistent (non-transient, even if often variable) objects in the Universe. They are bright in the entire electromagnetic spectrum. Blazars are a special class where the jets point nearly to our line of sight. Because of this special geometry and the bulk relativistic motion of the plasma in the jet, their radiation is enhanced by relativistic beaming. The majority of extragalactic objects detected in $\gamma$-rays are blazars. However, finding their counterparts in other wavebands could be challenging. Here we present the results of our 5-GHz European VLBI Network (EVN) observation of the radio source J1331+2932, a candidate blazar found while searching for possible $\gamma$-ray emission from the stellar binary system DG CVn (Loh et al. 2017). The highest-resolution radio interferometric measurements provide the ultimate tool to confirm the blazar nature of a radio source by imaging compact radio jet structure with Doppler-boosted radio emission, and give the most accurate celestial coordinates as well.}

\FullConference{14th European VLBI Network Symposium \& Users Meeting (EVN 2018)\\
		8-11 October 2018\\
		Granada, Spain}

\begin{document}

\section{Introduction}

When looking for evidence for $\gamma$-ray flaring activity in the stellar binary system DG CVn, recently Loh et al. \cite{Loh17} found a transient source using \emph{Fermi} Large Area Telescope (LAT) data from 2012 November. However, since simultaneous flaring of DG CVn was not reported at any other waveband that time, the background quasar J1331+2932 fell under suspicion as the possible source of the $\gamma$-rays. Among the $\gamma$-ray emitting extragalactic sources, blazars constitute the most populous group (Ackermann et al. 2015). Blazars are active galactic nuclei (AGN) with relativistic plasma jets directed at small inclination angles to the observer. Milliarcsecond (mas) resolution radio interferometric observations using the technique of very long baseline interferometry (VLBI) are the best suited for providing the ultimate evidence to discriminate between blazar and non-blazar radio-emitting AGN.

\section{Observations and data reduction}

We observed J1331+2932 with the European VLBI Network (EVN) at 5~GHz on 2017 Apr 11 (project code: RSG08, PI: K. \'E. Gab\'anyi). Nine telescopes of the EVN were used in e-VLBI mode: Jodrell Bank Mk2 (United Kingdom), Westerbork (the Netherlands), Medicina, Noto (Italy), Toru\'n (Poland), Yebes (Spain), Hartebeesthoek (South Africa), Irbene (Latvia), and Tianma (China). The long intercontinental baselines from the European stations to Hartebeesthoek and Tianma provided high angular resolution ($\sim 1.5$~mas) in both north--south and east--west directions. The maximum data rate was 2048~Mbit\,s$^{-1}$ but four antennas (Jodrell Bank, Westerbork, Toru\'n, and Tianma) operated at half of that value. The corresponding total bandwidth of 256~MHz was divided into 8 intermediate frequency (IF) channels in both right and left circular polarizations. Each IF was further divided into 64 spectral channels. Since the brightness and compactness of the source was previously unknown, the method of phase referencing was applied. This involved regular observations of the nearby ($1.4^\circ$ angular separation) compact calibrator J1334+3044 within the atmospheric coherence time, with a duty cycle of $\sim$6.5~min. The total time spent on the target source J1331+2932 was about 140 min. 

The EVN data were calibrated in the NRAO Astronomical Image Processing System (AIPS) \cite{Greisen03} and hybrid mapping was performed in Difmap \cite{Shepherd97} according to standard procedures. More details will be published elsewhere (A. Gemes et al., in prep.). The phase-referenced image of J1331+2932 was used to determine its accurate astrometric position with respect to the calibrator source. Our EVN observation prove that the target source was sufficiently bright and compact for direct fringe-fitting in AIPS. Therefore we made our final image of J1331+2932 this way.

\section{Results and discussion}

\begin{figure}
\center
	\includegraphics[width=0.5\columnwidth]{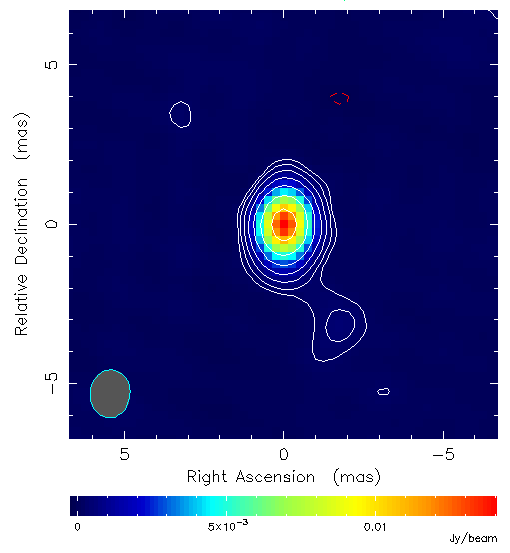}
	\caption{5-GHz EVN image of J1331+2932. The peak brightness is 14.08~mJy~beam$^{-1}$. The first contours are at $\pm 0.165$~mJy~beam$^{-1}$ ($\sim 3\sigma$ image noise), the positive contour levels increase by a factor of 2. The Gaussian restoring beam shown in the lower-left corner is 1.5~mas $\times$ 1.3~mas (FWHM) with a major axis position angle $-5.5^\circ$.}
	\label{image}
\end{figure}

The coordinates of the blazar candidate could be determined more accurately than before: right ascension 13$^{\rm h}$ 31$^{\rm m}$ 01.83259$^{\rm s}$ and declination 29$^\circ$ 32$^\prime$ 16.5099$^{\prime\prime}$. The estimated error in the position is 0.5~mas. The 5-GHz EVN image of J1331+2932 using fringe-fitted data is displayed in Fig.~\ref{image}. The source shows a compact radio structure typical for blazars, a bright core and a weak jet component to the south-west.

We fitted circular Gaussian brightness distribution model components to the self-calibrated visibility data in Difmap. For the core, the component shrank to a point regardless of whether a second Gaussian component was fitted to the jet or not. It implies that the compact core is practically unresolved in this experiment. We derived an upper limit of the size of the core (0.09~mas) by taking the minimum resolvable size \cite{Kovalev05} into account. This allowed us to estimate a lower limit to its brightness temperature \cite{Condon82}:
\begin{equation}
T_{\rm B}=\frac{2\ln{2}}{\pi}\frac{c^{2}S}{k_{B}\nu^{2}\vartheta^{2}}(1+z),
\end{equation}
where $c$ is the speed of light, $k_{B}$ the Boltzmann constant, $z=0.48$ the redshift of J1331+2932 \cite{Alam15}, $S$ the core flux density (14.7~mJy), $\nu$ the observing frequency, and $\vartheta$ the full width at half maximum (FWHM) angular size of the source (an upper limit in our case). This yields that the minimum brightness temperature of J1331+2932 is $T_B\geq1.3\times10^{11}$~K. Assuming an intrinsic brightness temperature $T_{\rm B,int}= 3 \times 10^{10}$~K \cite{Homan06}, the lower limit of the Doppler factor is $\delta = T_{\rm B} / T_{\rm B,int} \geq 4.3$. This indicates relativistic beaming which implies that the candidate is indeed a blazar. Using typical values of $5 \leq \Gamma \leq 15$ for the Lorentz factor, the jet inclination angle $\theta$ can be estimated using
\begin{equation}
\delta=\frac{1}{\Gamma\left(1-\beta\cos{\theta}\right)},
\end{equation}
where $\beta$ is the bulk speed of the jet in units of the speed of light. From this, we obtain $\theta \leq 14^{\circ}$ for the jet in J1331+2932.

Based on our EVN data, we confirm that J1331+2932 is a blazar and thus the most likely counterpart of the \emph{Fermi} LAT source. We also investigated mid-infrared monitoring data taken by the \emph{Wide-field Infrared Survey Explorer (WISE)} satellite \cite{Wright10}. These show significant flux density variations from daily to yearly time scales, strengthening the case for J1331+2932 being a blazar. Moreover, the \emph{WISE} colours of the source are close to typical values observed for blazars \cite{Massaro11,Massaro12}.

\end{document}